\documentclass[pre,preprint,english,showpacs,aps,prb,a4paper,floatfix]{revtex4}
\usepackage[T1]{fontenc}
\usepackage{color}
\usepackage[latin1]{inputenc}
\usepackage{graphicx}
\usepackage{bm}
\usepackage{epsfig}

\begin{document}
\title{Interplay between columnar and smectic stability in suspensions of polydisperse colloidal platelets}

\author{Enrique Velasco}
\email{enrique.velasco@uam.es}
\address{Departamento de F\'{\i}sica Te\'orica de la Materia Condensada,
Instituto de Ciencia de Materiales Nicol\'as Cabrera and Condensed Matter Physics Center (IFIMAC),
Universidad Aut\'onoma de Madrid, E-28049 Madrid, Spain.}

\author{Yuri Mart\'{\i}nez-Rat\'on}
\email{yuri@math.uc3m.es}
\address{Grupo Interdisciplinar de Sistemas Complejos (GISC),
Departamento de Matem\'{a}ticas,Escuela Polit\'{e}cnica Superior,
Universidad Carlos III de Madrid, Avenida de la Universidad 30, E--28911, Legan\'{e}s, Madrid, Spain}

\begin{abstract}
The phase behavior of a model suspension of colloidal polydisperse platelets is studied using density-functional theory.
Platelets are modelled as parallel rectangular prisms of square section $l$ and height $h$, with length and height
distributions given by different polydispersities $\delta_l$ and $\delta_h$. 
We obtain the phase behavior of the model, including nematic, smectic and columnar phases and its
dependence with the two polydispersities $\delta_l$ and $\delta_h$. When $\delta_l>\delta_h$ we observe that the smectic 
phase stabilises first with respect to the columnar. If $\delta_h>\delta_l$ we observe the opposite behavior. Other more 
complicated cases occur, e.g. the smectic stabilises from the nematic first but then exists a first-order transition to 
the columnar phase. Our model assumes plate-rod symmetry, but the regions of stability of smectic and columnar phases 
are non-symmetric in the $\delta_l-\delta_h$ plane due to the different dimensionality of ordering in the two phases. 
Microsegregation effects, i.e. different spatial distribution for different sizes within the periodic cell, take place 
in both phases.  
\end{abstract}

\pacs{61.30Cz, 61.30.Eb, 64.70.M, 47.57.J-}

\maketitle

\section{Introduction}

The issue of polydispersity is crucial to understand the phase behaviour of colloidal
suspensions of anisotropic particles, since
shape and size polydispersities have a profound impact on the phase behavior of colloidal liquid-crystal suspensions.
Near monodisperse colloids made of rod or plate-shaped particles present the usual cascade of liquid-crystal phase transitions
as the total volume fraction is increased: isotropic (I)-nematic (N)-smectic/columnar (S/C)-crystal (K).
However, colloidal particles can never be made truly identical in size and shape. Polydispersity gives rise to a complex phase 
behavior, with the presence of multiple phase coexistence between phases with different orientational and/or positional ordering
\cite{Kooij0,Kooij1,Vroege,Verhoeff,Leferink,Kooij2}.
Coexistence gaps are usually broadened and, for high polydispersities, strong demixing and fractionation are usually observed,
with coexisting phases having dissimilar size/shape distributions. Also new phenomena, such as
density inversion, are exclusive of polydisperse systems. In this case the more disordered phase (I)
becomes denser than the N phase \cite{Kooij2}. Polydispersity has also a dramatic impact on the
kinetic behavior of colloidal suspensions; a most remarkable effect consists of the extremely long times
necessary for the system to reach thermodynamic equilibrium \cite{Leferink}.

Models for polydisperse fluids of anisotropic particles should produce as an output the size or shape density distribution 
function $\rho({\bm r},\hat{\boldsymbol{\Omega}},\boldsymbol{\sigma})$, where ${\bm r}$ and $\hat{\bm\Omega}$ denote the spatial 
and angular particle degrees of freedom, while $\boldsymbol{\sigma}$ refers to the set of polydisperse
variables. The theoretical modelling of polydisperse fluids constitutes a complicated task due to the large number
of degrees of freedom involved in the calculations. This in turn translates into a numerical implementation of
the model which involves the evaluation of multiple integrals in a high-dimensional space. For this
reason, theories of polydisperse systems are formulated in terms of simplified models which postulate that
the excess part of the free-energy depends on a finite set of moments
of the density distribution function \cite{Warren,Sollich}. In this way the number of degrees of freedom are conveniently 
reduced and the problem becomes tractable. Within these models, the I-N \cite{Sluckin,Clarke,Sollich2} or N-N \cite{Sollich3}
equilibria of length-polydisperse freely-rotating rods were calculated. Also, within the restricted-orientation approximation,
the effect of polydispersity on the stability of the biaxial nematic phase in a mixture of uniaxial
rods and plates \cite{Yuri1,Yuri2} or in a one-component fluid of biaxial board-like particles
was recently studied \cite{Roij}.
The scarce MC simulation results on polydisperse anisotropic particles confirm the high fractionation
between the coexisting I and N phases of polydisperse infinitely-thin platelets \cite{Frenkel1}.
They also reveal the existence of a terminal polydispersity
beyond which the S phase of length-polydisperse hard rods becomes unstable with respect to the C phase \cite{Frenkel2}.

Polydispersity in size crucially affects the formation of phases
with spatial order, since it is difficult to accommodate the unit-cell dimension with the varying particle size.
Once stabilised, colloidal suspensions made of discs or platelets tend to form a N phase which changes to a
C phase as the particle volume fraction is increased. In the C phase particles stack one on top of the other to form
columns that in turn arrange in a two-dimensional lattice. An increasing polydispersity in lateral size (disc diameter)
tends to destabilize the C phase with respect to the N or S phases. In fact, above a certain threshold
value, the C phase turns into a S phase provided the polydispersity in thickness is not too large.
A large value of the latter discourages the formation of the layered S phase. In the case of suspensions of zirconium-phosphate
mineral plate-like particles \cite{Sue}, where the particle thickness is constant but the particle diameter is polydisperse, 
the S phase was found to be stable. The effect of polydispersity on the phase behavior of these suspensions was recently studied
from experimental and theoretical points of view \cite{Mejia,Kike1,Kike2}. It would be desirable to be able to rationalise
the effect of the different particle polydispersity coefficients on the phase behaviour, and to know in advance the ranges of 
values of the polydispersities where one can expect to find a particular phase.

\begin{figure}[h]
\includegraphics[width=7 cm]{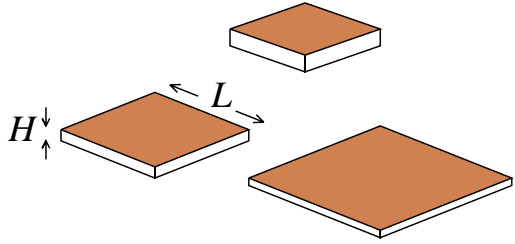}
\caption{\label{Fig1} \normalsize Schematic of square cuboids used in this work, with 
particle size parameters $L$ (side length) and $H$ (thickness) indicated.}
\end{figure}

In the present work we study the effect that the size polydispersity has on the relative stability between
the liquid-crystal phases with partial positional ordering, in particular between the S and C phases.
We use density-functional theory to analyse a suspension of colloidal
platelets made from hard square cuboids, i.e. rectangular prisms of square cross section $L$
and thickness $H$, Fig. \ref{Fig1}.
The normal axes of the cuboids are taken to point along a common
direction and rotations about this direction are not allowed, the sides of the
particles being always parallel. This approximation does not appear to be too unrealistic considering that the orientational
order parameter will be high close to the transition from the N phase to either the S or C phases.
This model can be analysed using a generalisation
of the density-functional theory for hard cubes derived
in Ref. \cite{Yuri3} and used for the first time to calculate the phase diagram of the one-component and binary mixture 
fluids \cite{Yuri4}. In contrast to the latter work, where bidisperse
cubes were considered, here we introduce a continuous distribution in both $L$ and $H$,
respectively characterized by the distribution variances $\delta_l$ and
$\delta_h$, with a view to obtaining phase diagrams involving the N, S and C phases
as a function of the two polydispersities and the particle volume fraction.

In Section \ref{SM} we review the density-functional theory used and give some details on the numerical methodology. 
Results are presented in Section \ref{Results},
and a short discussion and the conclusions are given in Section \ref{conclusions}.

\section{Model and numerical methodology}
\label{SM}

\noindent{\it Density functional}. 
The density-functional theory used is based on the fundamental-measure formalism for mixtures of parallel hard cubes. 
This formalism was derived by Cuesta and Mart\'{\i}nez-Rat\'on \cite{Yuri1}, and here
we generalise it to general polydisperse fluids. The excess free-energy
functional  in units of thermal energy $kT=\beta^{-1}$, is $\beta {\cal F}_{\rm ex}[\rho]=
\int_V d{\bm r} \Phi({\bm r})$, where
\begin{eqnarray}
\Phi&=&
-n_0\log{(1-n_3)}+
\frac{{\bm n}_1\cdot{\bm n}_2}{1-n_3}+\frac{n_{2x}n_{2y}n_{2z}}{(1-n_3)^2},
\label{1}
\end{eqnarray}
and where $n_{\alpha}$ are average densities,
\begin{eqnarray}
n_{\alpha} ({\bm r})=\!\int_0^{\infty}\!\!\! dL \int_0^{\infty}\!\!\!dH \int_V d {\bm r}^{\prime}
\rho({\bm r}^{\prime};L,H) \omega^{(\alpha)}({\bm r}^{\prime}-{\bm r};L,H).
\label{2}
\end{eqnarray}
The index $\alpha$ takes the values $\{0,1x,1y,1z,2x,2y,2z,3\}$.
$\omega^{(\alpha)}$ are the particle geometrical measures \cite{Yuri1}.
$\rho({\bm r};L,H)$ is the local density of
particles with lengths $L$ and $H$ at the point ${\bm r}$. We define $l=L/L_0$ and $h=H/H_0$, with $L_0=\left<L\right>$ and 
$H_0=\left<H\right>$ the mean length and height, respectively. In the following we take $L_0$ as a unit of length, i.e. $L_0=1$.
Due to the parallel particle approximation, the physics of the problem scales in the $z$ direction, so that we can also
take $H_0=1$. In fact, the following equivalence takes place: ${\cal F}[\rho;L_0,H_0]v_0\equiv 
{\cal F}[\rho^*;1,1]$, where $\rho^*({\bm r},l,h)=\rho({\bm r},l,h) L_0 H_0 v_0$ 
(with $v_0=L_0^2H_0$ the mean volume of particle) is the dimensionless 
density distribution function. The difference between both free-energies, coming from the ideal part, is proportional to the total number 
of particles and does not affect the phase behavior of the system.  
This scaling implies that the same functional 
can be used to describe a fluid of parallel platelets and a fluid of parallel rods. This equivalence will be used 
later to compare with different experimental results on polydisperse platelets and rods. The total free energy is then
\begin{eqnarray}
\beta{\cal F}[\rho]= \int_0^{\infty} dl\int_0^{\infty} dh \int_V d{\bm r}
\rho({\bm r};l,h)\left\{\log{\left[\rho({\bm r};l,h)\Lambda^3(l,h)\right]}-1\right\}
+\beta {\cal F}_{\rm ex}[\rho],
\label{free}
\end{eqnarray}
which has to be minimised with respect to $\rho({\bm r};l,h)$ for each phase.
$\Lambda(l,h)$ is the thermal wavelength for particles with size $(l,h)$, which in principle can be
adsorbed into the chemical potential of that species.
Here we consider the N phase, where $\rho({\bm r};l,h)=\rho(l,h)$,
the S phase, with $\rho({\bm r};l,h)=\rho(z;l,h)$, and the C
phase, where $\rho({\bm r};l,h)=\rho({\bm r}_{\perp};l,h)$, and
${\bm r}_{\perp}=(x,y)$. The common particle axis is taken along $\hat{\bm z}$.\\

An important aspect of the problem is the polydispersity of the parent solution.
We can write $\rho({\bm r};l,h)=\rho_0 f({\bm r};l,h)$, where $\rho_0=N/V$ is the total mean density, $N$ the
total number of particles, $V$ the volume, and the function $f({\bm r};l,h)$ is the
local fraction of particles of species $(l,h)$. It satisfies:
\begin{eqnarray}
\frac{1}{V}\int_V d{\bm r}\int_0^{\infty} dl\int_0^{\infty} dh f({\bm r};l,h)=1.
\label{6}
\end{eqnarray}
The global fraction is
\begin{eqnarray}
x(l,h)=\frac{1}{V}\int_Vd{\bm r} f({\bm r};l,h)=\frac{1}{N}\int_Vd{\bm r}\rho({\bm r};l,h),
\end{eqnarray}
and obviously
\begin{eqnarray}
\int_0^{\infty} dl\int_0^{\infty} dh x(l,h)=1.
\end{eqnarray}
We assume that $x(l,h)$, the size distribution of the parent solution, is fixed and can be factorised (i.e. length and thickness
distributions may be assumed to be uncorrelated; in the real world this may be correct or not in depending on the 
particle synthesis methodology).
Therefore we make the following assumption:
\begin{eqnarray}
x(l,h)=\phi(l)\phi(h),
\end{eqnarray}
%where $\phi(l)$ and $\phi_h(h)$ are length and height distributions. We take
where $\phi(\sigma)$; $\sigma=l$ or $h$ depending on whether one refers to the length or thickness distribution. We take
\begin{eqnarray}
\phi(\sigma)=D\sigma^{\gamma}e^{-\lambda \sigma^2}
\end{eqnarray}
in terms of two parameters $\gamma$ and $\lambda$.
We impose the condition that the first two moments are equal to unity, which fix the
normalisation and the mean value
$\left<x\right>=1$. Then:
\begin{eqnarray}
&&1=\int_0^{\infty}d\sigma
\phi(\sigma)\hspace{0.4cm}\rightarrow\hspace{0.4cm}D=\frac{2\lambda^{(\gamma+1)/2}}
{\displaystyle\Gamma\left(\frac{\gamma+1}{2}\right)},\nonumber\\\nonumber\\
&&1=\int_0^{\infty}d\sigma \sigma\phi(\sigma)\hspace{0.4cm}\rightarrow\hspace{0.4cm}
\lambda=\left[\frac{\displaystyle\Gamma\left(\frac{\gamma+2}{2}\right)}{\displaystyle\Gamma\left(\frac{\gamma+1}{2}\right)}\right]^2.
\end{eqnarray}
Then the second moment can be related uniquely to the polydispersity coefficient:
\begin{eqnarray}
\left<\sigma^2\right>&=\int_0^{\infty}d\sigma \sigma^2\phi(\sigma)=\frac{\displaystyle\Gamma\left(\frac{\gamma+1}{2}\right)\Gamma\left(\frac{\gamma+3}{2}\right)}
{\displaystyle\Gamma\left(\frac{\gamma+2}{2}\right)^2},
\label{l22}
\end{eqnarray}
and the polydispersity coefficient $\delta$ is:
\begin{eqnarray}
\delta^2=\frac{\left<\sigma^2\right>- \left<\sigma\right>^2}{\left<\sigma\right>^2}=\left<\sigma^2\right>-1.
\label{delta_gamma}
\end{eqnarray}
From here, the $\gamma$ parameter can be obtained given an input value for $\delta$. The equation must be
solved numerically. Note that, once we fix the values of $\delta_l$ and $\delta_h$, the values of $\gamma_l$ 
and $\gamma_h$ are in general different.

Using the free-energy functional (\ref{free}), the equilibrium condition reads:
\begin{eqnarray}
\beta\mu(l,h)&=&\frac{\delta\beta{\cal F}}{\delta \rho({\bm r};l,h)}=
\log{\left[\rho({\bm r};l,h)\Lambda^3(l,h)\right]}-c^{(1)}({\bm r};l,h),
\nonumber\\\nonumber\\&\rightarrow&\rho({\bm r};l,h)=e^{\displaystyle\beta\mu^*(l,h)}e^{\displaystyle c^{(1)}({\bm r};l,h)},
\label{4}
\end{eqnarray}
where $\mu(l,h)$ is the chemical potential of the species $(l,h)$, with
\begin{eqnarray}
\mu^*(l,h)=\mu(l,h)-3\log{\Lambda(l,h)},
\end{eqnarray}
and $c^{(1)}({\bm r};l,h)$ the
one-body direct correlation function. Integrating Eqn. (\ref{4}) and eliminating $\mu^*(l,h)$, 
the Euler-Lagrange equation to be solved is:
\begin{eqnarray}
f({\bm r};l,h)=x(l,h)\frac{\displaystyle e^{\displaystyle c^{(1)}({\bm r};l,h)}}
{\displaystyle\frac{1}{V}\int_Vd{\bm r} e^{\displaystyle c^{(1)}({\bm r};l,h)}}.
\label{7}
\end{eqnarray}
We define the moments
\begin{eqnarray}
m_{\lambda\eta}({\bm r})= \int_0^{\infty}dl l^{\lambda}\int_0^{\infty} dh h^{\eta} 
f({\bm r};l,h).
\end{eqnarray}
Multiplying (\ref{7}) by $l^{\lambda}h^{\eta}$, where $\lambda,\eta$ are integers, 
and integrating over all possible values of $l$ and $h$:
\begin{eqnarray}
m_{\lambda\eta}({\bm r})=\int_0^{\infty}dl l^{\lambda}\int_0^{\infty} dh h^{\eta}
x(l,h)&\left[\frac{\displaystyle e^{\displaystyle c^{(1)}({\bm r};l,h)}}
{\displaystyle\frac{1}{V}\int_Vd{\bm r} e^{\displaystyle c^{(1)}({\bm r};l,h)}}\right].
\label{9}
\end{eqnarray}
If we can express $c^{(1)}$ in terms of the moments, Eqns. (\ref{9}) form a closed set of
equations for the moments.

Since there is no dependence on spatial coordinates in the 
N phase, Eqns. (\ref{1}),(\ref{2}) and (\ref{free}) can be used directly to evaluate the free energy.
In the case of S and C phases, we will deal with the spatial dependence by using Fourier
representations. Because of the particular mathematical structure of the density functional, the one-body
direct correlation function can be expressed in terms of the average densities $n_{\alpha}({\bm r})$ only.
Therefore it is possible to use Fourier expansions for these local densities and reconstruct the local
fraction of particles $f({\bm r};l,h)$ using (\ref{7}) evaluated at the equilibrium average densities.
In the S phase, the unknown coefficients will be $c_{\alpha}^{(k)}$, with
\begin{eqnarray}
n_{\alpha}(z)=\rho_0\sum_{k=0}^{\infty}c_{\alpha}^{(k)}\cos(qkz),
\label{n1}
\end{eqnarray}
and $q=2\pi/d$ the wavevector associated with the S period $d$. In the C phase, the 
unknown coefficients will be $\Upsilon_{\alpha}^{(nm)}$, with
\begin{eqnarray}
&&n_{\alpha}({\bm r}_{\perp})= 
\rho_0\sum_{n,m=0}^{\infty}\Upsilon_{\alpha}^{(nm)}\cos(qnx)\cos(qmy),
\label{n2}
\end{eqnarray}
where $q=2\pi/a$ and $a$ the lattice parameter of the square lattice. 
The Fourier expansions in (\ref{n1}) and (\ref{n2}) 
were truncated so that the absolute value of the highest-order coefficient included was less that $10^{-6}$ for all the 
densities explored. In each case the Euler-Lagrange equation is solved by iterations until convergence of the coefficients. 
Appendix \ref{app2} provides more details on the numerical methodology.

A first picture of the order in which the different phases appear is provided by
a bifurcation analysis. Here one perturbs the N phase with a small-amplitude density wave of given wavevector ${\bm q}$ and 
searches for instability with respect to density and wavelength. Instability is given by the curvature of the free-energy
functional, expressed by the direct correlation functional 
$c^{(2)}({\bm r},{\bm r}^{\prime},l,l',h,h')=-\beta\delta^2{\cal F}_{\hbox{\tiny ex}}[\rho]/\delta\rho({\bm r},l,h)
\delta\rho({\bm r}',l',h')$. The bifurcation 
point (density at which the N becomes unstable with respect to S- or C-like perturbations) coincides with 
the transition point whenever the true phase transition is continuous; otherwise the bifurcation point only provides the
spinodal point and a full treatment based on equality of pressure and partial chemical potentials is needed.
We provide details on the bifurcation analysis in Appendix \ref{bifur}.
   
\section{Results}
\label{Results}
\begin{figure}[h]
\includegraphics[width=6.5in,angle=0]{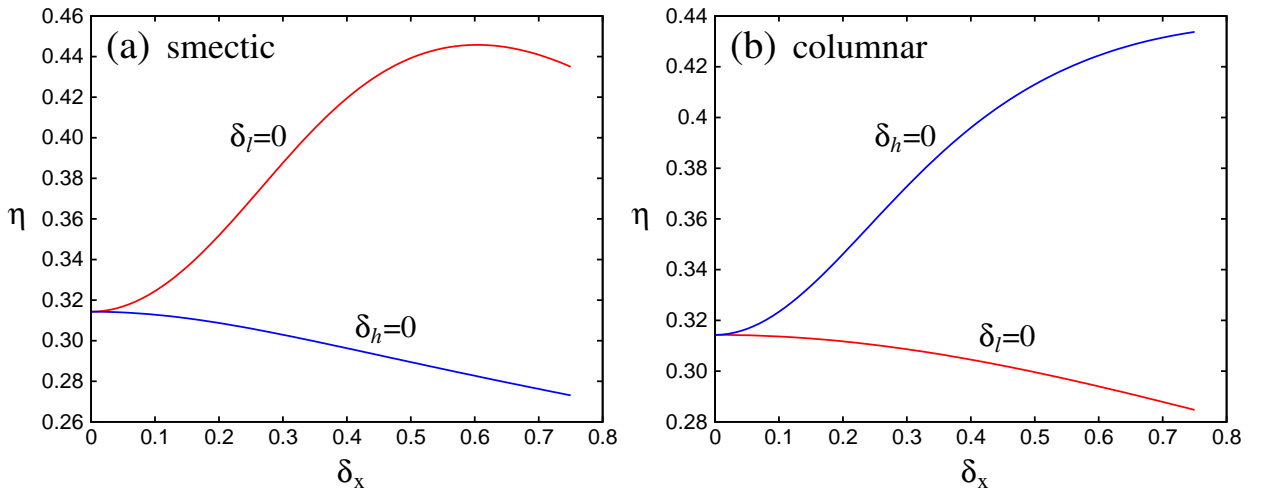}
\caption{Packing fractions $\eta$ at which the N phase bifurcates to the (a) S, (b) C phases, as a function of
one of the polydispersities $\delta_x$ ($x=l, h$) when the other polydispersity is set to zero.}
\label{bif}
\end{figure}

To explore the overall structure of phase behaviour, we have first computed the bifurcation densities, i.e. the densities
at which the Euler-Lagrange equations have a spatially inhomogeneous solution (bifurcation from the N to the S or
C phases). In the following we will use the packing fraction $\eta$, defined by $\eta=\rho_0\left<L^2\right>\left<H\right>=
\rho_0\left(1+\delta_l^2\right)$, as the relevant density parameter. Fig. \ref{bif} shows the bifurcation 
packing fractions $\eta$ for the S and C phases as a function of one of the polydispersities 
when the other polydispersity is set to zero.

In the case of bifurcation to the S phase, panel (a), we see that, when $\delta_l=0$ (red curve) and the thickness
polydispersity $\delta_h$ increases, the packing fraction also does due to the increasing difficulty to create uniform
layers in the system. The decrease in $\eta$ at large values of $\delta_h$ is a microsegregation effect.
Here the thickness distribution of particles is different in the layers and in between the layers (interstitials); 
in the former, the thickness distribution is peaked about a larger value of thickness than in the interstitials.  
We will comment on this effect later. The opposite case is when $\delta_h=0$ (blue curve). Here,
as $\delta_l$ increases, the S bifurcates at lower packing fractions since platelets can better pack in the
(quasi two-dimensional) layers when their side-length distribution is wider.

The bifurcation to the C phase, Fig. \ref{bif}(b), is relatively similar to the S case, except that the
roles of $l$ and $h$ are interchanged. The most notable difference is that the maximum in $\eta$ does not exist, although
there could also be a microsegregation effect where small platelets are expelled from the nodes of the columns to the
interstitial space, as discussed later.  

\begin{figure}[h]
\includegraphics[width=5.5in,angle=0]{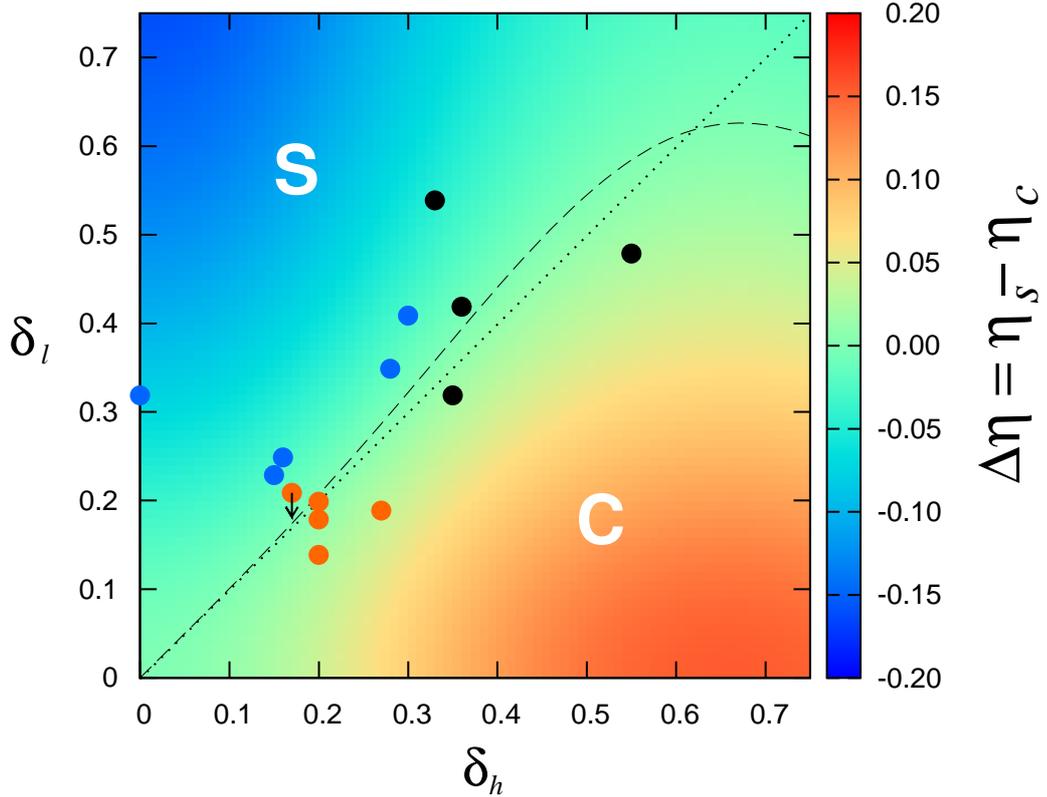}
\caption{Difference $\Delta\eta = \eta_s -\eta_c$ between the packing fractions of the bifurcated S and
C phases as a function of polydispersity coefficients $\delta_l$ and $\delta_h$, in false colour.
The dashed line corresponds to the curve $\Delta\eta=0$ where the S and C phases bifurcate at the same value of
packing fraction. The S (C) label indicates the region where the S (C) phase bifurcates from the N first. 
Dotted line is the bisectrice $\delta_l=\delta_h$. Orange, blue and black filled circles correspond to experimental 
values of polydispersities where C, S and C+S were found. In the case of black circles and the orange circle with a down arrow, 
polydispersities correspond to those of the parent phase.}
\label{sabana}
\end{figure}

The overall effect of polydispersities can be visualised in the plot of Fig. \ref{sabana}, where the difference in 
bifurcation packing fractions $\Delta\eta = \eta_s -\eta_c$ for the S, $\eta_s$, and C, $\eta_c$, phases
is shown as a function of the polydispersity coefficients $\delta_l$ and $\delta_h$. The curve $\Delta\eta=0$ indicates
a situation where both phases bifurcate at the same packing fraction (projected black dashed curve in the figure). We see that,
for large values of $\delta_l$ and $\delta_h$, this curve does not correspond or is close to the bisectrice $\delta_l=\delta_h$ 
(the dotted black curve).
This means that the C phase stands a higher value for its associated polydispersity coefficient ($\delta_l$) than
does the S ($\delta_h$) up to $\delta_h\approx 0.6$; for higher values, the scenario is the opposite.

\begin{figure*}[h]
\includegraphics[width=6.2in,angle=0]{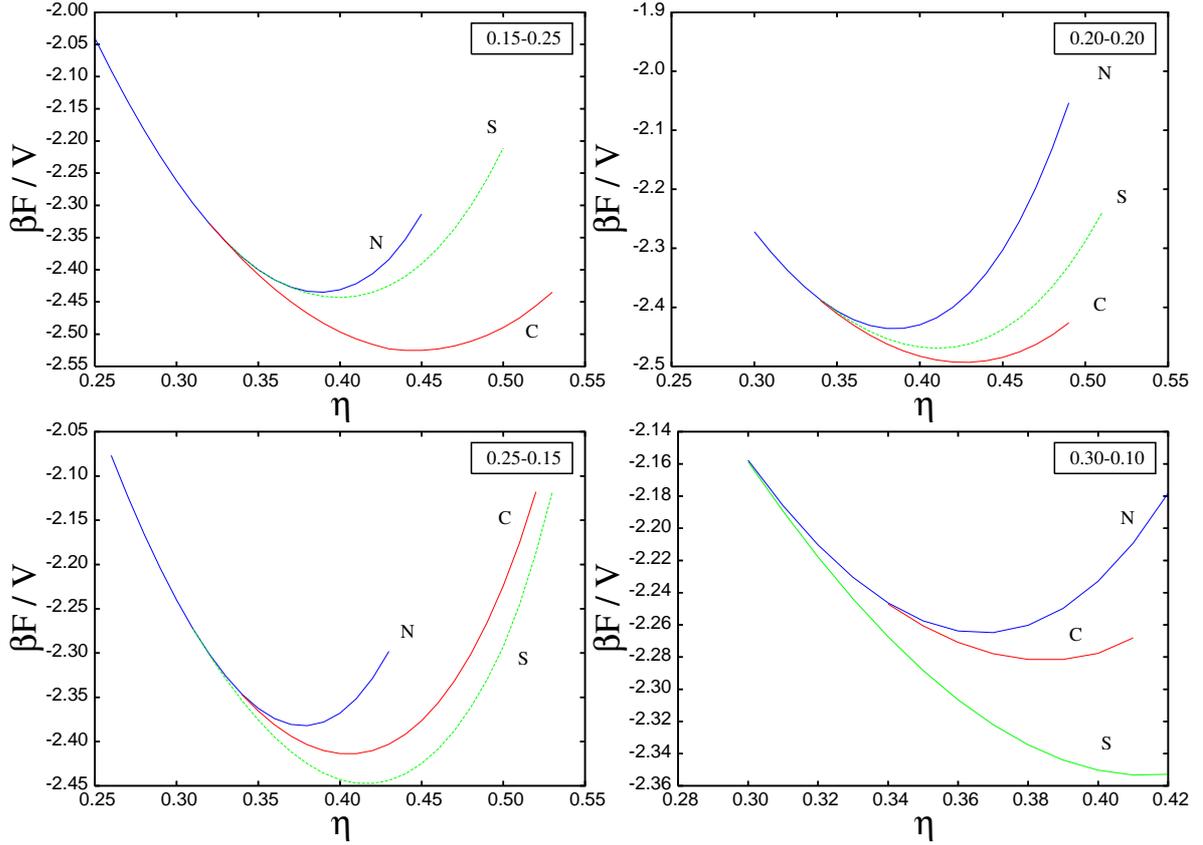}
\caption{Free-energy density per unit of thermal energy, $\beta{\cal F}/V$, as a function of platelet packing fraction $\eta$,
for different values of the polydispersities $(\delta_l,\delta_h)$. (a) (0.15,0.25), (b) (0.20,0.20), (c) (0.25,0.15), 
(d) (0.30,0.10). Free-energy branches for N, S and C phases are represented in blue, green and red, respectively, and
labelled by the corresponding symbol.} 
\label{frees}
\end{figure*}

A more detailed treatment of the problem requires a full analysis beyond linear bifurcation theory, i.e. a full
solution to the Euler-Lagrange equation (\ref{7}). This allows to obtain the free energy and the local density
distribution of the different phases. 
Fig. \ref{frees} shows some representative cases. 
In this series of graphs we are following the line $\delta_l+\delta_h=0.4$ in the $\delta_l$--$\delta_h$ plane.
The reference case is a completely monodisperse solution, 
$(\delta_l,\delta_h)=(0.0,0.0)$, for which S and C phases bifurcate at the same density, but the C
phase is always more stable. Panel (a) refers to the case $(\delta_l,\delta_h)=(0.10,0.30)$. Here the C phase
bifurcates before the S, and is much more stable than the S; this is because the thickness polydispersity is
relatively high and prevents the formation of regular layers. As $\delta_l$ increases and $\delta_h$ decreases, the stability
gap between the two phases decreases, panels (b) and (c), and for the case $(0.25,0.15)$ the S phase is already more
stable and bifurcates before the C phase. All of these results corroborate the results of the bifurcation analysis
presented before, in the sense that a high polydispersity in one parameter ($\delta_l$ or $\delta_h$) inhibits the formation of 
the corresponding phase (C or S, respectively). They also point to the fact that 
the $\delta_l$ and $\delta_h$ parameters do not play a symmetric role: the C phase stands a
higher value of polydispersity in its associated polydispersity than the S, and when $\delta_l=\delta_h$ the C phase 
is more stable than the S.

Note that there are some cases where the S phases bifurcates before and remains more stable than the C 
phase but only below some density [this occurs for values between those of panels (c) and (d)]; at higher 
densities the two branches should cross each other and the C phase becomes more stable (the phase transition 
between the two phases is of first order although no attempt has been made to calculate the properties of 
the coexisting phases).

\begin{figure}[h]
\includegraphics[width=6.2in,angle=0]{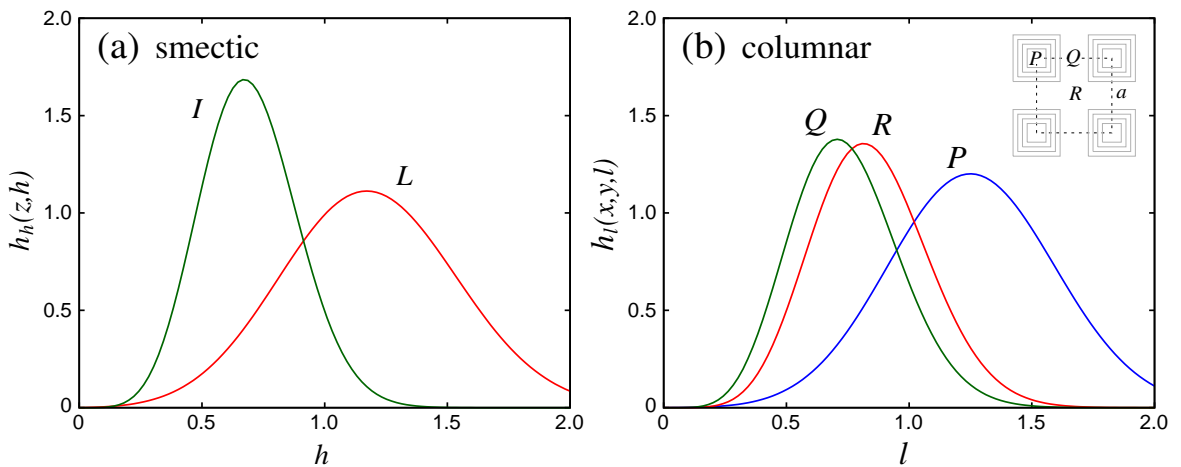}
\caption{(a) The function $h_h(z,h)$ defined in Eqn. (\ref{hh}) for a C phase $\delta_l=0.10$, $\delta_h=0.20$ at
packing fraction $\eta=0.40$. The function is evaluated at two points, L (layer, $z=0$) and I (interstitial, $z=d/2$, i.e.
half S period). (b) The function $h_l({\bm r}_{\perp},l)$ defined in Eqn. (\ref{hh}) for a C phase $\delta_l=0.20$, $\delta_h=0.10$ at
packing fraction $\eta=0.416$. The function is evaluated at three points in the square-lattice unit cell, $P$, $R$ and $Q$, 
which are indicated in the inset ($P$ is a lattice site, $R$ the interstitial point, and $Q$ a point midway between two
neighboring lattice sites).}
\label{struc1}
\end{figure}

The structure of the S and C phases is also interesting to investigate due to the fact that we expect strong
microsegregation effects in these phases caused by the spatial ordering. By this we mean that the particle size distribution 
will depend on the location within the periodic unit cell. To better visualise the microsegregation effect occurring at the 
scale of the periodic unit cell in the S phase, we define the function
\begin{eqnarray}
h_h(z,h)=\frac{\displaystyle\int_0^{\infty}dl f(z,l,h)}{\displaystyle\int_0^{\infty}dl\int_0^{\infty} dh 
f(z,l,h)},
\label{hh0}
\end{eqnarray}
which gives the particle thickness $h$ distribution at point $z$ irrespective of the side-length $l$.
Fig. \ref{struc1}(a) plots $h_h(z,h)$ as a function of $h$ for the case $\delta_l=0.10$, $\delta_h=0.20$ and at the points
located at the S layers and at the interstitial point (midway between two consecutive layers).  
We can see that the first is mostly populated by the thickest particles, while the interstitial mostly 
contains the thinner ones. Note that, as usual, the moment $m_{00}({\bm r})$ is higher at the lattice sites that at the 
interstitials.

As for the C phase, to quantify the microsegregation effect occurring at the scale of the periodic unit cell, we
define the function
\begin{eqnarray}
h_l({\bm r}_{\perp},l)=\frac{\displaystyle\int_0^{\infty}dh f({\bm r}_{\perp},l,h)}
{\displaystyle\int_0^{\infty}dl\int_0^{\infty} dh 
f({\bm r}_{\perp},l,h)},
\label{hh}
\end{eqnarray}
which gives the particle size-length distribution at point ${\bm r}_{\perp}$ irrespective of the thickness. In Fig.
\ref{struc1}(b) we illustrate the microsegregation effect by plotting the distribution $h_l({\bm r}_{\perp},l)$ at three points 
of the square-lattice unit cell: P, at the nodes; R, at the interstitials; and
Q, midway between two nodes. The curves indicate that big particles tend to occupy the nodes, while small particles
are more probable in the interstitials. 

Up to now there have appeared in the literature a number of
well-controlled experiments on suspensions of colloidal platelets. There are two problems when comparing the theory with
the experiment. One is that polydispersities are difficult to measure with precision, especially the polydispersity in
thickness. Normally one assumes that $\delta_h$ is some positive and constant value (particle synthesis normally involves
exfoliation and the thickness usually consists of one or a few sheets of the original layered material), subject to large
error, while $\delta_l$ is measured much more accurately and can be finely tuned more easily. The other is that
suspensions are subject to gravity and the effect of sedimentation and phase profile is crucial to understand the
true phase behaviour of the suspension \cite{Dani}. Usually this effect is not properly taken into account.

A number of groups have obtained suspensions made of particles with interactions that can be approximated as hard interactions.
The Dutch group have been profusely investigating suspensions of 
gibbsite particles with different polydispersities $\delta_l$. Their $\delta_h$ is quite large and somewhat uncontrolled,
so that their suspensions usually exhibit a N-to-C transition. On the other hand, the A\&M Texas group have been
synthesising mineral platelets with $\delta_h=0$ (although the interactions may not always be considered as hard); in these
systems the tendency to form S phases is quite strong. In a recent paper, the latter group have provided quantitative
data on polydispersity in $\delta_l$ in their samples \cite{Sue}.
 
In Fig. \ref{sabana} some of the above experimental results on colloidal suspensions of mineral particles have been added, specially
those that are relevant to elucidate the relative stability of the C and S phases. Whenever available, we plot
the polydispersity coefficients $(\delta_h,\delta_l)$ as measured in the fractionated coexisting phases; note that in some
experiments the (in general different) polydispersities of the two coexisting phases are not measured, but only the parent 
or global polydispersities. In the figure, orange circles show the values of the coefficients corresponding to experiments on
gibbsite platelet suspensions where stable C \cite{Kooij0}, hexatic C \cite{hexatic}
and hexagonal C \cite{hexagonal} phases were found. The arrow pointing down in one of the orange circles
means that the values of $(\delta_h,\delta_l)$ correspond to the parent phase. We expect that, after fractionation,
$\delta_l$ will decrease. Black points correspond to polydispersities of the parent distribution
for suspensions of goethite nanorods \cite{chem_matt,Pol}, which phase separate into C and S phases. Finally, blue circles 
with $\delta_h\neq 0$ represent values for which stable
S \cite{Pol} or smectic B \cite{smectic_B} phases were found in suspensions of goethite nanorods \cite{Pol} or of charged
colloidal gibbsite platelets \cite{smectic_B}. The blue circle with $\delta_h=0$ is a result from recent experiments on
Zirconium-Phosphate mineral platelet suspensions of constant thickness but polydisperse in diameter \cite{Sue}. As can be seen,
the values of $(\delta_h,\delta_l)$ corresponding to colloidal suspensions where the C (orange) or S (blue) phases were
found to be stable are approximately located in their regions predicted by our stability calculations.

\section{Conclusions}
\label{conclusions}

In this paper we have investigated the effect of the thickness and width polydispersities on the relative stability between
the S and C phases in a system of aligned board-like uniaxial particles, using a density-functional theory for hard cubes extended 
to a polydisperse mixture. An understanding on the effect of polydispersity begins with the phase behaviour of the corresponding
one-component (monodisperse) system. At high densities, the system is in a K phase but the free-energy difference with the C
phase is very small \cite{PHC1,PHC2}. The S phase is much less stable. All phases bifurcate from the same 
point (the corresponding packing fraction being $\eta^*=0.3143$). On inclusion of polydispersity
in both particle thickness and width, the K phase is expected to destabilise more strongly because the three-dimensional
periodic arrangement of particles in the unit cell is very sensitive to the increase of polydispersity as compared to the
higher-symmetry C and S phases. This is the reason why we have not included the K phase in the present study.

We have modelled the polydispersities in a symmetric way: the mean thickness and side length are fixed to unity, and the 
global size distribution is a product of two independent functions which have the same functional form in their 
respective size variable. Therefore the
mean particle geometry is cubic and the addition of both polydispersities deforms the original geometry to be prolate or oblate 
as the side length and thickness polydispersity coefficients $\delta_l$ and $\delta_h$ become nonzero.
We have made a bifurcation analysis to study the effect of polydispersity on the S and C bifurcation from the N phase. 
The difference $\Delta \eta=\eta_{\rm s}^*-\eta^*_{\rm c}$ between bifurcation packing fractions allowed us to conclude that, in 
a first approximation, for 
relatively low values of $\delta_l$ and $\delta_h$ and for $\delta_l>\delta_h$ the S phase destabilizes with respect to the N 
phase first, while the C phase destabilizes first when $\delta_l<\delta_h$. The curve $\Delta \eta=0$ in the 
$\delta_l-\delta_h$ plane is located close to the bisectrice for $\delta_h<0.6$ (but slightly favouring the 
stability of the C phase), 
a result due to the symmetric way in which particle polydispersities are included. However, for $\delta_h>0.6$, 
the curve $\Delta\eta$ deviates from the bisectrice and favours the S phase. A possible explanation for this lies in the
microfractionation mechanism: for large polydispersities larger particles tend to occupy the sites of the periodic
lattice (whether one dimensional in the case of the S phase or two-dimensional for the C phase) and small particles tend 
to occupy the lattice interstitials; this mechanism may be more effective in the S phase than in the C phase.
In fact, experiments show \cite{Kooij0} that the C phase accepts a polydispersity of at least $\delta_l=0.25$, but
recent studies on length-polydisperse rods of goethite \cite{Pol} indicate that the S phase is stable for polydispersities as 
large as $\delta_h=0.55$ (with due allowance for macrofractionation mechanisms which reduce the polydispersity of the parent
sample).

The bifurcation analysis does not give the final answer to the question of relative phase behavior. To have a more
profound understanding of the phase diagram of the present system and investigate the microsegregation effects, we have 
performed numerical minimizations of the free-energy density functional of our polydisperse fluid, fixing the 
probability size distribution of the parent phase $x(l,h)$ (which coincides with the unit-cell distribution in a periodic phase). 
We have found that, in general, the bifurcation analysis gives the correct relative stability of the C and S phases. 
However, there are some values of polydispersities around the $\delta_l=\delta_h$ bisectrice for which the 
S phase bifurcates from the N before the C phase, but at some value of packing fraction the C and S free-energy branches
cross, the C phase becoming energetically favoured at higher packing fractions. 

This behaviour is a clear indication that the system should exhibit a first order S-C transition.
To correctly predict the particle composition of the coexisting phases, cloud-shadow
coexistence calculations should be performed in which the coexistence size distributions $x_{c,s}(l,h)$ are
calculated from equations expressing the equality of chemical potentials of all species and the pressures,
together with the level rule (conservation of the number of particles) \cite{Sollich}.
These calculations imply a heavy numerical task since the moments of the coexisting distributions
involve the evaluation of multiple integrals and many non-linear integral equations have to be solved.
Therefore, in the present paper we restricted our effort to the calculation of free-energy branches.

Finally, to confirm the microsegregation scenario, we examined the actual structure of the phases and the spatial distribution 
of species with different sizes. In effect, the layers or sites are preferentially populated by big particles, while the 
interstitials have a higher proportion  of small particles. Particles with intermediate sizes have a similar composition in both locations,
indicating their relatively high diffusivity. For high polydispersities, this extra degree of freedom conspires with the 
different dimensionality of ordering in the two phases (one or two in the S and C phases, respectively) to give a non-symmetric 
stability map in a fluid of our otherwise perfectly symmetric particles.

\acknowledgments

Financial support from Comunidad Aut\'onoma de Madrid (Spain) under
the R$\&$D Programme of Activities MODELICO-CM/S2009ESP-1691, and from MINECO (Spain)
under grants FIS2010-22047-C01 and FIS2010-22047-C04 is acknowledged.

\appendix
\section{Numerical details}
\label{app2}

Here we give more details about the way we performed the numerical calculations.\\
\\
\noindent{\it Smectic phase}. The
Euler-Lagrange equation for the local fraction of particles is
\begin{eqnarray}
f(z;l,h)=\frac{\displaystyle x(l,h) e^{\displaystyle \Delta c^{(1)}(z;l,h)}}
{\displaystyle\frac{1}{d}\int_0^d dz' e^{\displaystyle \Delta c^{(1)}(z';l,h)}},
\label{a1}
\end{eqnarray}
where $d$ is the S period, and $\Delta c^{(1)}(z;l,h)=c^{(1)}(z;l,h)-c_0^{(1)}(l,h)$, with 
$c_0^{(1)}(l,h)$ the excess part of the chemical potential (in thermal units $KT$) or bulk one-body 
direct correlation function (this is done to improve numerical accuracy).
Now we expand in Fourier space:
\begin{eqnarray}
f(z;l,h)=\sum_{k=0}^{\infty}f_k(l,h)\cos(kqz),
\end{eqnarray}
where $q=2\pi/d$ and $f_0(l,h)=1$. Multiplying (\ref{a1}) by a cosine function
and integrating over one period, we obtain
\begin{eqnarray}
f_k(l,h)=2x(l,h)\frac{\displaystyle{\int_0^d dz e^{\displaystyle \Delta c^{(1)}(z;l,h)}\cos(kqz)}}
{\displaystyle\int_0^d dz e^{\displaystyle \Delta c^{(1)}(z;l,h)}},
\label{EL3}
\end{eqnarray}
with $k>0$. Now using the definitions
\begin{eqnarray}
&&\hspace{-0.8cm}c_{rs}^{(k)}\equiv\int_0^{\infty}dl l^{r}\int_0^{\infty}dh h^{s} f_k(l,h)\cos{\left(\frac{kqh}{2}\right)},
\nonumber\\\nonumber\\
&&\hspace{-0.8cm}s_{rs}^{(k)}\equiv\frac{2}{kq}\int_0^{\infty}dl l^{r}\int_0^{\infty}dh h^{s} f_k(l,h)
\sin{\left(\frac{kqh}{2}\right)},
\label{coeffff}
\end{eqnarray}
with $r,s=0,1,...$,
the corresponding projections of the 
Euler-Lagrange equations, given by (\ref{EL3}), can be rewritten solely in terms of the $\{c_{rs}^{(k)},s_{rs}^{(k)}\}$
coefficients. This is because, as can be easily shown, the average functions $n_{\alpha}$ depend on these coefficients only:
\begin{eqnarray}
n_{\alpha}(z)=\rho_0\sum_{k=0}^{\infty}d_{\alpha}^{(k)}\cos(qkz),
%\label{n1}
\end{eqnarray}
where, for each value of $k>0$,
\begin{eqnarray}
d_{\alpha}=\{c_{00},c_{10},c_{10},s_{00},s_{10},s_{10},c_{20},s_{20}\}
\end{eqnarray}
for $\alpha=\{0,1x,1y,1z,2x,2y,2z,3\}$ respectively.
This means that the $c^{(1)}$ correlation function can be uniquely expressed in terms of these coefficients and, consequently,
Eqns. (\ref{EL3}) and (\ref{coeffff}) form a closed set of equations in these coefficients. To solve these equations we
use an iterative method with a mixing parameter of $0.5$. The solution is obtained after typically 30 iterations, and starting
values for the coefficients were chosen from the values obtained for a previous density.

\noindent{\it Columnar phase}. In this case the Euler-Lagrange equation reads
\begin{eqnarray}
f({\bm r}_{\perp};l,h)=\frac{\displaystyle x(l,h)e^{\displaystyle \Delta c^{(1)}({\bm r}_{\perp};l,h)}}
{\displaystyle\frac{1}{a_c}\int_{a_c}d{\bm r}_{\perp} e^{\displaystyle \Delta c^{(1)}({\bm r}_{\perp};l,h)}},
\label{A6}
\end{eqnarray}
where $a_c=a^2$ is the area of the unit cell of the square lattice, and $a$ the lattice parameter, and 
$\Delta c^{(1)}({\bm r}_{\perp};l,h)=c^{(1)}({\bm r}_{\perp};l,h)-c_0^{(1)}(l,h)$. Expanding in a
Fourier series,
\begin{eqnarray}
f({\bm r}_{\perp};l,h)=\sum_{n,m=0}^{\infty} f_{nm}(l,h)\cos(qnx)\cos(qmy),
\label{aqui}
\end{eqnarray}
with $f_{00}(l,h)=1$, and $q=2\pi/a$. Multiplying Eqn. (\ref{aqui}) by cosine functions in $x$ and $y$ and 
integrating over the unit cell, we obtain: 
\begin{eqnarray}
f_{nm}(l,h)=(1+\delta_{n0})(1+\delta_{m0}) x(l,h)
\frac{\displaystyle{\int_{a_c}d{\bm r}_{\perp} e^{\displaystyle{\Delta c^{(1)}({\bm r}_{\perp};l,h)}}\cos(qnx)\cos(qmy)}}
{\displaystyle{\int_{a_c}d{\bm r}_{\perp} e^{\displaystyle{\Delta c^{(1)}({\bm r}_{\perp};l,h)}}}}.
\end{eqnarray}
Using the definition
\begin{eqnarray}
\Upsilon_{rs,uv}^{(nm)}\equiv\int_0^{\infty}\!\!\!\! dll^u\int_0^{\infty}\!\!\!\! dhh^{v}f_{nm}(l,h)
\Xi_{r}\!\!\left(\frac{nql}{2}\right)\Xi_{s}\!\!\left(\frac{mql}{2}\right),
\end{eqnarray}
with $r,s=0,1$, $u,v=0,1,...$, and $\Xi_{0}(x)=\cos{x}$, $\Xi_{1}(x)=\sin{x}/x$, Eqn. (\ref{A6}) can be written as a
transcendental set of equations in $\{\Upsilon_{rs,uv}^{(nm)}\}$, which form a closed set of equations because the average
densities can be written in terms of solely these coefficients,
\begin{eqnarray}
n_{\alpha}({\bm r}_{\perp})=\rho_0\sum_{n,m=0}^{\infty}e_{\alpha}^{(nm)}\cos(qnx)\cos(qmx),
\end{eqnarray}
with 
\begin{eqnarray}
e_{\alpha}=\{\Upsilon_{00,00},\Upsilon_{10,10},\Upsilon_{01,10},\Upsilon_{00,01},\Upsilon_{01,11},\Upsilon_{10,11},\Upsilon_{11,20},\Upsilon_{11,21}\},
\end{eqnarray}
for $\alpha=\{0,1x,1y,1z,2x,2y,2z,3\}$ respectively, and for each value of $n,m$. An iterative method was used to solve 
the equations with a mixing parameter of $0.5$ and typically $100$ iterations were necessary to reach convergence.
Starting values for a given density were obtained from the solution of a previous density.  

\section{Bifurcation analysis}
\label{bifur}
In this section we give details on the bifurcation analysis from the N phase with respect to the S or C phases of 
the polydisperse mixture.
Eqn. (\ref{4}) can be rewritten as
\begin{eqnarray}
\rho({\bm r};l,h)=\rho_0 x(l,h) e^{\displaystyle{\Delta c^{(1)}({\bm r};l,h)}},
\label{bif1}
\end{eqnarray}
The nonuniform phase bifurcates from
the parent phase and, near the bifurcation point, we
can assume that $\rho({\bm r};l,h)\approx \rho_0 x(l,h)+\epsilon({\bm r};l,h)$
where $\epsilon({\bm r};l,h)$ is an small perturbation. Inserting this
into Eqn. (\ref{bif1}) and functionally expanding the exponential up to
first order in $\epsilon({\bm r};l,h)$, we arrive at
\begin{eqnarray}
\epsilon({\bm r};l,h)=\rho_0 x(l,h)\int_0^{\infty} dl'\int_0^{\infty} dh'c^{(2)}({\bm r}-{\bm r}';
l',h')\epsilon({\bm r}';l',h'),
\label{eps}
\end{eqnarray}
with $c^{(2)}({\bm r}-{\bm r}'; l,h,l',h')$ the direct correlation function: 
\begin{eqnarray}
\left.c^{(2)}({\bm r}-{\bm r}';l,h,l',h')=
\frac{\delta c^{(1)}({\bm r};l,h)}{\delta \rho({\bf r}',l',h')}\right|_{\rho({\bm r};l,h)=\rho_0 x(l,h)}.
\end{eqnarray}
In Fourier space, Eqn. (\ref{eps}) becomes
\begin{eqnarray}
\hat{\epsilon}({\bm q};l,h)=\rho_0x(l,h)\int_0^{\infty} dl'\int_0^{\infty} dh'\hat{c}^{(2)}
({\bm q};l',h')\hat{\epsilon}({\bm q};l',h'),
\label{epsilon}
\end{eqnarray}
where, as usual, $\hat{f}({\bm q})=\int d{\bm r} e^{i{\bm q}\cdot {\bm r}}
f({\bm r})$. The Fourier transform of the direct correlation
function can be written as
\begin{eqnarray}
-\hat{c}^{(2)}({\bm q};l',h')=\sum_{\alpha,\beta}
\Phi_{\alpha\beta}(\rho_0)\hat{\omega}^{(\alpha)}
({\bm q};l',h')\hat{\omega}^{(\beta)}
({\bm q};l',h'),
\label{corre}
\end{eqnarray}
where the coefficients $\displaystyle{
\Phi_{\alpha\beta}=\frac{\partial^2\Phi}{\partial n_{\alpha}
\partial n_{\beta}}}$ are evaluated at $\rho({\bm r};l,h)=\rho_0x(l,h)$.
Therefore these coefficients are functions of $\rho_0$ and $\delta_l$ since, in the uniform limit, the weighted densities 
$n_{\alpha}({\bm r})$ are
$n_0=n_{1x}=n_{1y}=n_{1z}=n_{2x}=n_{2y}=\rho_0$ and 
$n_{2z}=n_3=\rho_0\left(1+\delta_l^2\right)$.   
$\hat{\omega}^{(\alpha)}({\bm q};l,h)$
are the Fourier transforms of the weights $\omega^{(\alpha)}({\bm r};l,h)$. Substitution
of Eqn. (\ref{corre}) into Eqn. (\ref{epsilon}) gives
\begin{eqnarray}
\hat{\epsilon}({\bm q};l,h)=-\rho_0x(l,h)\sum_{\alpha,\beta}\hat{\omega}^{(\alpha)}({\bm q};l,h)
\Phi_{\alpha\beta}(\rho_0)s_{\beta}({\bm q}),
\label{inter}
\end{eqnarray}
where we have defined
\begin{eqnarray}
s_{\beta}({\bm q})=\int_0^{\infty} dl\int_0^{\infty} dh \hat{\omega}^{(\beta)}
({\bm q};l,h)\hat{\epsilon}({\bm q};l,h).
\end{eqnarray}
Multiplying (\ref{inter}) by $\hat{\omega}^{(\tau)}({\bm q};l,h)$ and
integrating over $l$ and $h$, we find
\begin{eqnarray}
{\bm s}({\bm q})=-\rho_0\left[\hat{T}(\rho_0,{\bm q})\cdot \hat{\Phi}(\rho_0)
\right]{\bm s}({\bm q}),
\label{find}
\end{eqnarray}
where ${\bm s}({\bm q})$ is a column vector of dimension eight with
coordinates $s_{\beta}({\bm q})$ ($\beta=0,2,1x,1y,1z,2z,2y,2z$),
$\hat{\Phi}(\rho_0)$ is the matrix with elements $\Phi_{\alpha\beta}$, and
$\hat{T}(\rho_0,{\bm q})$ is the $8\times 8$ matrix with
elements
\begin{eqnarray}
T_{\alpha\beta}=\int_0^{\infty} dl\int_0^{\infty} dh x(l,h)\hat{\omega}^{(\alpha)}
({\bm q};l,h)\hat{\omega}^{(\beta)}({\bm q};l,h).
\label{last}
\end{eqnarray}
Defining the matrix
\begin{eqnarray}
\hat{H}(\rho_0,{\bm q})=I+\rho_0\hat{T}(\rho_0,{\bm q})\cdot
\hat{\Phi},
\end{eqnarray}
with $I$ the $8\times 8$ identity matrix, Eqn. (\ref{find})
can be rewritten as
\begin{eqnarray}
\hat{H}(\rho_0,{\bm q}){\bm s}({\bm q})=0.
\label{nontrivial}
\end{eqnarray}
Thus, a nontrivial solution of (\ref{nontrivial}) can be calculated from
\begin{eqnarray}
{\cal H}(\rho_0,{\bm q})=0,\hspace{0.6cm}\boldsymbol{\nabla}_{{\bm q}}
{\cal H}(\rho_0,{\bm q})=0,
\label{final}
\end{eqnarray}
i.e. by searching for the equality to zero of the global minimum
of ${\cal H}(\rho_0,{\bm q})=\text{det}\left[\hat{H}(\rho_0,{\bm q})\right]$.
The bifurcation to the S or C phases can be obtained using
${\bm q}=(0,0,q)$ or ${\bm q}=(q,0,0)$, respectively, and the solutions of
(\ref{final}) furnish the values of density and wavevector, $\rho_0^*$ and $q^*$, at bifurcation.
Now taking into account the factorised form of the weighting functions,  
\begin{eqnarray}
\hat{\omega}^{(\alpha)}({\bm q};l,h)=\hat{\omega}^{(\alpha)}_x(q_x;l)\hat{\omega}^{(\alpha)}_y(q_y;l)
\hat{\omega}^{(\alpha)}_z(q_z;h),
\end{eqnarray}
and of the parent size probability distribution $x(l,h)=\phi(l)\phi(h)$, the 
coefficients (\ref{last}) can be written as a product of two one-dimensional integrals:
\begin{eqnarray}
T_{\alpha\beta}({\bm q})=\left[\int_0^{\infty}dl\phi(l)\prod_{\tau=\alpha,\beta}\hat{\omega}^{(\tau)}_x(q_x;l)
\hat{\omega}^{(\tau)}_y(q_y;l)\right] 
\times
\left[\int_0^{\infty}dh\phi(h)\prod_{\tau=\alpha,\beta}\hat{\omega}^{(\tau)}_z(q_z;h)\right].
\end{eqnarray}
Finally, substituting the values $q_x=q_y=0,q_z=q$ or $q_x=q,q_y=q_z=0$ in these integrals, we find that they can be 
expressed as a function of the following integrals:
\begin{eqnarray}
S_n(q)&=&\int_0^{\infty}d\sigma \phi(\sigma)\sigma^n \sin(q\sigma),\nonumber\\
C_n(q)&=&\int_0^{\infty}d\sigma \phi(\sigma)\sigma^n \cos(q\sigma),
\end{eqnarray}
where $\sigma=\{l,h\}$. In turn, these integrals can efficiently be calculated as
\begin{eqnarray}
S_n(q)&=&q\langle\sigma^{n+1}\rangle \exp\left(-\frac{q^2}{4\lambda}\right)
M\left[-\frac{(\gamma+n-1)}{2},\frac{3}{2},\frac{q^2}{4\lambda}\right],\nonumber\\\nonumber\\
C_n(q)&=&\langle\sigma^n\rangle \exp\left(-\frac{q^2}{4\lambda}\right)
M\left[-\frac{(\gamma+n)}{2},\frac{1}{2},\frac{q^2}{4\lambda}\right],
\end{eqnarray}
where $M[a,b,x]$ is the Confluent Hypergeometric function of real 
arguments (the so-called {\it Kummer function} \cite{Gradshteyn}), while the general 
expression for the $n$-th moment of the distribution function 
$\phi(\sigma)$ is
\begin{eqnarray}
\langle\sigma^n\rangle=\frac{\Gamma\left[\displaystyle{\frac{\gamma+n+1}{2}}\right]
\Gamma^{n-1}\left[\displaystyle{\frac{\gamma+1}{2}}\right]}
{\Gamma^n\left[\displaystyle{\frac{\gamma+2}{2}}\right]}.
\end{eqnarray}

\end{document}